\documentclass[twocolumn,prb]{revtex4}
\usepackage{graphicx}
\usepackage{dcolumn}
\usepackage{bm}
\usepackage{amsmath}

\begin{document}

\preprint{}
\title{Manifestation of the odd-frequency spin-triplet pairing state in
diffusive ferromagnet / superconductor junctions}
\author{T. Yokoyama$^1$, Y. Tanaka$^1$ and A. A. Golubov$^2$}
\affiliation{$^1$Department of Applied Physics, Nagoya University, Nagoya, 464-8603, Japan%
\\
and CREST, Japan Science and Technology Corporation (JST) Nagoya, 464-8603,
Japan \\
$^2$ Faculty of Science and Technology, University of Twente, 7500 AE,
Enschede, The Netherlands}
\date{\today}

\begin{abstract}
Using the quasiclassical Green's function formalism, we study the
influence of the odd-frequency spin-triplet superconductivity on the
local density of states (LDOS) in a diffusive ferromagnet (DF)
attached to a superconductor. Various possible symmetry classes in a
superconductor are considered which are consistent with the Pauli's
principle: even-frequency spin-singlet even-parity (ESE) state,
even-frequency spin-triplet odd-parity (ETO) state, odd-frequency
spin-triplet even-parity (OTE) state and odd-frequency spin-singlet
odd-parity (OSO) state. For each of these states, the pairing state
in DF is studied. Particular attention is paid to the study of
spin-singlet $s$-wave and spin-triplet $p$-wave superconductors as
the examples of ESE and ETO superconductors. For spin-singlet case
the magnitude of the OTE component of the pair amplitude is enhanced
with the increase of the exchange field in DF. When the OTE
component is dominant at low energy, the resulting LDOS in DF has a
zero energy peak (ZEP). On the other hand, in DF / spin-triplet
$p$-wave superconductor junctions LDOS has a ZEP in the absence of
the exchange field, where only the OTE paring state exists. With the
increase of the exchange field, the ESE component of the pair
amplitude induced in DF is enhanced. Then, the resulting LDOS has a
ZEP splitting. We demonstrate that the appearance of the dominant
OTE component of the pair amplitude is the physical reason of the
emergence of the ZEP of LDOS.
\end{abstract}

\maketitle



%

%




\section{Introduction}

Ferromagnet/superconductor structures with conventional spin-singlet
$s$-wave superconductors have been the subject of extensive work
during the past decade \cite{buzdinrev,Efetov2,golubovrev}. An exciting
manifestation of anomalous proximity effect in these structures is
the existence of the so-called $\pi$-junctions in SFS Josephson
junctions confirmed
experimentally in \cite%
{Ryazanov,Kontos,Se03,Blu04,Su02,Be02,She06,Wei05,Se04}. Recently, diffusive
ferromagnet/superconductor (DF/S) junctions have received much attention due to
the possibility of generation of the odd-frequency pairing in these
structures \cite{Efetov1,Efetov2}. In DF, due to the isotropization by the
impurity scattering, only even-parity $s$-wave pairing is allowed. Besides
this, the exchange field breaks the time reversal symmetry and both
spin-singlet and spin-triplet Cooper pairs can coexist. In accordance with
the Pauli's principle, this spin-triplet state belongs to the odd-frequency
spin-triplet even-parity (OTE) pairing \cite{Efetov1,Efetov2}. Various
aspects of this state have been addressed in recent theoretical work \cite%
{Efetov2,Efetov3,Kulic,Kadigro,Eschrig,Fominov} and first experimental
observation of the long-range proximity effect due to the odd-frequency
pairing was reported in \cite{Kaizer,Sosnin}.

Odd-frequency pairing is an unique state which was first proposed by
Berezinskii \cite{Berezinskii} as a hypothetical state of $^{3}$He. The
odd-frequency superconductivity was then discussed in the context of various
pairing mechanisms involving strong correlations \cite{Balatsky,Vojta,Fuseya}%
. However, proximity effect in the presence of odd-frequency superconducting
state has not been studied up to very recently.

A general theory of the proximity effect in junctions composed of diffusive
normal metal (DN) and unconventional superconductor in the framework of the
quasiclassical Green's function formalism was recently presented by two of
the present authors \cite{Tanaka2006}. Various possible symmetry classes in
a superconductor were considered \ in Ref.\cite{Tanaka2006} which are
consistent with the Pauli's principle: even-frequency spin-singlet
even-parity (ESE) state, even-frequency spin-triplet odd-parity (ETO) state,
odd-frequency spin-triplet even-parity (OTE) state and odd-frequency
spin-singlet odd-parity (OSO) state. For each of the above four cases,
symmetry and spectral properties of the induced pair amplitude in the DN
were determined. It was shown that the pair amplitude in a DN belongs
respectively to ESE, OTE, OTE and ESE pairing states. It is remarkable that
OTE state is realized without assuming magnetic ordering in DN/ETO
superconductor junctions, where the mid gap Andreev resonant state \cite%
{Tanaka} formed at the interface penetrates into the DN and the resulting
local density of states (LDOS) has a zero energy peak (ZEP) \cite{p-wave}.

On the other hand, the existence of ZEP in LDOS in the DF/ ESE $s$-wave
superconductor junctions has been established \cite%
{Buzdin1,Kontos,Zareyan,Bladie,Bergeret1,golubov}. Although the conditions
of the formation of ZEP in DF regions were formulated by the present authors
\cite{Yoko}, possible relation between the ZEP and the formation of OTE
paring in DF has not been yet clarified. The present paper addresses this
issue. We also study the proximity effect in DF/ETO $p$-wave superconductor
junctions. It was shown in the previous paper \cite{Tanaka2006} that only
the OTE pairing state is generated without exchange field $h$. It is an
interesting question how this unusual proximity effect is influenced by the
exchange field.

The organization of the present paper is as follows. In section II, we
formulate the proximity effect model in DF / S junctions within the theory
applicable to unconventional superconductor junctions where the MARS are
naturally taken into account in the boundary condition for the
quasiclassical Green's function \cite{p-wave}. We discuss the general
properties of the proximity effect by choosing ESE, ETO, OTE, and OSO
superconductor junctions. It is clarified that the OTE, ESE, ESE and OTE
states are, respectively, generated in the DF in the presence of exchange field $h$. In
section III we calculate the pair amplitude in DF for spin-singlet $s$-wave
and spin-triplet $p$-wave superconductor junctions as an example of ESE and
ETO superconductor junctions. For $s$-wave junctions, it is revealed that a
generation of the OTE pairing state by the exchange field $h$ causes an
enhancement of the zero energy LDOS in the DF. On the other hand, for $p$%
-wave superconductor junctions, a generation of ESE pairing state by $h$
results in a splitting of ZEP of LDOS. We clarify the relation between the
ZEP in LDOS and the generation of the OTE state in the DF. The summary of
the results is given in section IV.

\section{Formulation}

Let us start with the formulation of the general symmetry properties of the
quasiclassical Green's functions in the considered system following the
discussion in the Ref. \onlinecite{Tanaka2006}. The elements of retarded and
advanced Nambu matrices $\widehat{g}^{R,A}$
\begin{equation}
\widehat{g}^{R,A}=\left(
\begin{array}{cc}
g^{R,A} & f^{R,A} \\
\tilde{f}^{R,A} & \tilde{g}^{R,A}%
\end{array}%
\right)
\end{equation}%
are composed of the normal $g_{\alpha ,\beta }^{R}(\bm{
r},\varepsilon ,\bm{p})$ and anomalous ${f}_{\alpha ,\beta }^{R}(\bm{ r}%
,\varepsilon ,\bm{p})$ components with spin indices $\alpha $ and $\beta $.
Here $\bm{p}=\bm{p}_{F}/\mid \bm{p}_{F}\mid $, $\bm{p}_{F}$ is the Fermi
momentum, $\bm{r}$ and $\varepsilon $ denote coordinate and energy of a
quasiparticle measured from the Fermi level respectively. The function $f^{R}
$ and the conjugated function $\tilde{f}^{R}$ satisfy the following relation
\cite{Serene,Eschrig1}
\begin{equation}
\tilde{f}_{\alpha ,\beta }^{R}(\bm{ r},\varepsilon ,\bm{p})=-[f_{\alpha
,\beta }^{R}(\bm{ r},-\varepsilon ,\bm{-p})]^{\ast }.
\end{equation}

The Pauli's principle is formulated in terms of the retarded and the
advanced Green's functions in the following way \cite{Serene}
\begin{equation}
f_{\alpha ,\beta }^{A}(\bm{ r},\varepsilon ,\bm{p})=-f_{\beta ,\alpha }^{R}(%
\bm{ r},-\varepsilon ,\bm{-p}).
\end{equation}%
By combining the above two equations, we obtain $\tilde{f}_{\beta ,\alpha
}^{R}(\bm{ r},\varepsilon ,\bm{p})=[f_{\alpha ,\beta }^{A}(\bm{ r}%
,\varepsilon ,\bm{p})]^{\ast }$. Further, the definitions of the
even-frequency and the odd-frequency pairing are $f_{\alpha ,\beta }^{A}(%
\bm{ r},\varepsilon ,\bm{p})=f_{\alpha ,\beta }^{R}(\bm{ r},-\varepsilon ,%
\bm{p})$ and $f_{\alpha ,\beta }^{A}(\bm{ r},\varepsilon ,\bm{p})=-f_{\alpha
,\beta }^{R}(\bm{ r},-\varepsilon ,\bm{p})$, respectively. Finally we get
\begin{equation}
\tilde{f}_{\beta ,\alpha }^{R}(\bm{ r},\varepsilon ,\bm{p})=[f_{\alpha
,\beta }^{R}(\bm{ r},-\varepsilon ,\bm{p})]^{\ast }  \label{Even}
\end{equation}%
for the even-frequency pairing and
\begin{equation}
\tilde{f}_{\beta ,\alpha }^{R}(\bm{ r},\varepsilon ,\bm{p})=-[f_{\alpha
,\beta }^{R}(\bm{ r},-\varepsilon ,\bm{p})]^{\ast }  \label{Odd}
\end{equation}%
for the odd-frequency pairing. In the following, we consider a homogeneous
ferromagnet/superconductor junctions with the exchange field $h$ in a
ferromagnet and focus on the Cooper pairs with $S_{z}=0$. In this case, it
is possible to remove the external phase of the pair potential in the
superconductor. We will concentrate on the retarded part of the Green's
function.

We consider a junction consisting of a normal (N) and a superconducting
reservoirs connected by a quasi-one-dimensional diffusive ferromagnet (DF)
with a length $L$ much larger than the mean free path as shown in Fig. 1.

\begin{figure}[htb]
\begin{center}
\scalebox{0.4}{
\includegraphics[width=20.0cm,clip]{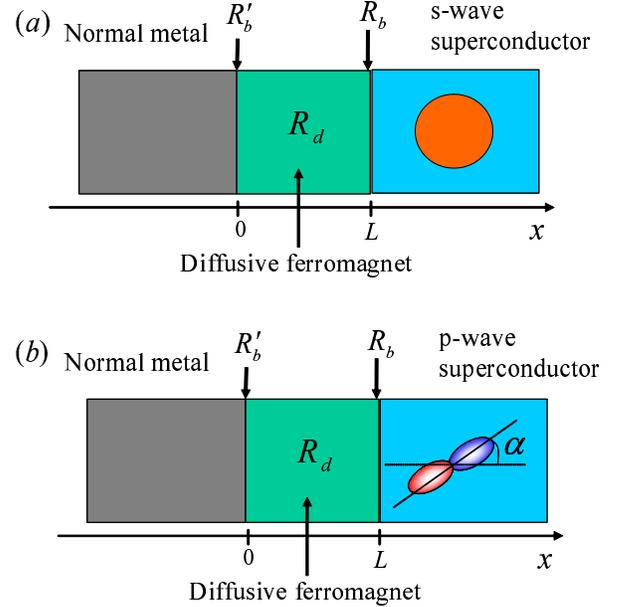}}
\end{center}
\caption{ (color online) Schematic illustration of DF /S junctions where DF
is connected to normal reservoirs. (a)conventional spin-singlet $s$-wave
superconductor and (b)spin-triplet $p$-wave superconductor junctions. }
\label{f1}
\end{figure}
The interface between the DF and the superconductor (S) at $x=L$ has a
resistance $R_{b}$ and the N/DF interface at $x=0$ has a resistance $%
R^{\prime }_{b}$. The Green's function in the superconductor can be
parameterized as $g_{\pm }(\varepsilon )\hat{\tau}_{3}+f_{\pm }(\varepsilon )%
\hat{\tau}_{2}$ using Pauli's matrices, where the subscript $+(-)$ denotes
the right (left) going quasiparticles. $g_{\pm }(\varepsilon )$ and $f_{\pm
}(\varepsilon )$ are given by $g_{+}(\varepsilon ) \equiv g_{\uparrow,
\uparrow }^{R}(\bm{ r},\varepsilon ,\bm{p}) = g_{\downarrow,\downarrow }^{R}(%
\bm{ r},\varepsilon ,\bm{p})$, $g_{-}(\varepsilon ) \equiv g_{\uparrow,
\uparrow }^{R}(\bm{ r},\varepsilon, \bar{\bm{p}}) = g_{\downarrow,\downarrow
}^{R}(\bm{ r},\varepsilon ,\bar{\bm{p}})$, $f_{+}(\varepsilon ) \equiv
f_{\uparrow,\downarrow}^{R}(\bm{ r},\varepsilon ,\bm{p})$, and $%
f_{-}(\varepsilon ) \equiv f_{\uparrow ,\downarrow }^{R}(\bm{ r},\varepsilon
,\bar{\bm{p}})$, respectively, with $\bar{\bm{p}}=\bar{\bm{p}}_{F}/\mid %
\bm{p}_{F}\mid$ and $\bar{\bm{p}}_{F}=(-p_{Fx},p_{Fy})$. Using the relations
(\ref{Even}) and (\ref{Odd}), we obtain that $f_{\pm }(\varepsilon )=[f_{\pm
}(-\varepsilon )]^{\ast }$ for the even-frequency pairing and $f_{\pm
}(\varepsilon )=-[f_{\pm }(-\varepsilon )]^{\ast }$ for the odd-frequency
pairing, respectively, while $g_{\pm }(\varepsilon )=[g_{\pm }(-\varepsilon
)]^{\ast }$ in both cases.

In the DF region, only the $s$-wave even-parity pairing state is allowed due
to isotropization by impurity scattering. The resulting Green's function
with majority and minority spin  in the DF can be
parameterized by $\cos \theta \hat{\tau}_{3}+\sin \theta \hat{\tau}_{2}$ and
$\cos \bar{\theta} \hat{\tau}_{3}+\sin \bar{\theta} \hat{\tau}_{2}$ in a
junction with an even-parity superconductor respectively. On the other hand,
for odd-parity superconductor, the corresponding quantities for majority
spin and minority spin are expressed by $\cos\theta \hat{\tau}_{3}+\sin
\theta \hat{\tau}_{1}$ and $\cos\bar{\theta} \hat{\tau}_{3}+\sin \bar{\theta}
\hat{\tau}_{1}$ respectively.

The function $\theta $ satisfies the Usadel equation \cite{Usadel}
\begin{equation}
D\frac{\partial ^{2}\theta }{\partial x^{2}}+2i(\varepsilon +h)\sin \theta =0
\label{eq.1}
\end{equation}%
with the boundary conditions at the DF/S interface \cite{p-wave,TGK}
\begin{equation}
\frac{L}{R_{d}}(\frac{\partial \theta }{\partial x})\mid _{x=L}=\frac{%
\langle F_{1}\rangle }{R_{b}},  \label{eq.2}
\end{equation}%
\begin{equation}
F_{1}=\frac{2T_{1}(f_{S}\cos \theta _{L}-g_{S}\sin \theta _{L})}{%
2-T_{1}+T_{1}(\cos \theta _{L}g_{S}+\sin \theta _{L}f_{S})}
\end{equation}%
and at the N/DF interface
\begin{equation}
\frac{L}{R_{d}}(\frac{\partial \theta }{\partial x})\mid _{x=0}=-\frac{%
\langle F_{2}\rangle }{R_{b}^{\prime }},F_{2}=\frac{2T_{2}\sin \theta _{0}}{%
2-T_{2}+T_{2}\cos \theta _{0}},  \label{eq.3}
\end{equation}%
respectively, with $\theta _{L}=\theta \mid _{x=L}$ and $\theta _{0}=\theta
\mid _{x=0}$. Here, $R_{d}$ and $D$ are the resistance and the diffusion
constant in the DF, respectively. Function $g_{S}$ is given by $%
g_{S}=(g_{+}+g_{-})/(1+g_{+}g_{-}+f_{+}f_{-})$ and $%
f_{S}=(f_{+}+f_{-})/(1+g_{+}g_{-}+f_{+}f_{-})$ for the even-parity pairing
and $f_{S}=i(f_{+}g_{-}-f_{-}g_{+})/(1+g_{+}g_{-}+f_{+}f_{-})$ for the
odd-parity pairing, respectively, with $g_{\pm }=\varepsilon /\sqrt{%
\varepsilon ^{2}-\Delta _{\pm }^{2}}$, $f_{\pm }=\Delta _{\pm }/\sqrt{\Delta
_{\pm }^{2}-\varepsilon ^{2}}$ and $\Delta _{\pm }=\Delta \Psi (\phi _{\pm })
$ where $\Psi (\phi _{\pm })$ is the form factor with $\phi _{+}=\phi $ and $%
\phi _{-}=\pi -\phi $. The brackets $\langle \ldots \rangle $ denote
averaging over the injection angle $\phi $:
\begin{equation}
\langle F_{1(2)}(\phi )\rangle =\int_{-\pi /2}^{\pi /2}d\phi \cos \phi
F_{1(2)}(\phi )/\int_{-\pi /2}^{\pi /2}d\phi T_{1(2)}\cos \phi ,
\label{average}
\end{equation}%
\begin{equation}
T_{1}=\frac{4\cos ^{2}\phi }{Z^{2}+4\cos ^{2}\phi },\;\;T_{2}=\frac{4\cos
^{2}\phi }{Z^{\prime }{}^{2}+4\cos ^{2}\phi },
\end{equation}%
where $T_{1,2}$ are the transmission probabilities, $Z$ and $Z^{\prime }$
are the barrier parameters for two interfaces.
%

The resistance at the interface $R_{b}^{(\prime )}$ is given by
\begin{equation*}
R_{b}^{(\prime )}=\frac{2R_{0}^{(\prime )}}{\int_{-\pi /2}^{\pi /2}d\phi
T_{1(2)}(\phi )\cos \phi }.
\end{equation*}%
Here, $R_{b}^{(\prime )}$ denotes $R_{b}$ or $R_{b}^{\prime }$, and $%
R_{0}^{(\prime )}$ is Sharvin resistance, which in three-dimensional case is
given by $R_{0}^{(\prime )}=4\pi ^{2})/(e^{2}k_{F}^{2}S_{c}^{(\prime )})$,
where $k_{F}$ is the Fermi wave-vector and $S_{c}^{(\prime )}$ is the
constriction area.

Next, we focus on the Green's function of minority spin. The function $\bar{%
\theta}$ satisfies the following equation \cite{Usadel}:

\begin{equation}
D\frac{\partial ^{2}\bar{\theta} }{\partial x^{2}} +2i(\varepsilon - h) \sin
\bar{\theta} =0  \label{eq.1n}
\end{equation}%
with the boundary condition at the DF/S interface \cite{p-wave,TGK}
\begin{equation}
\frac{L}{R_{d}}(\frac{\partial \bar{\theta} }{\partial x})\mid _{x=L}=\frac{%
\langle \bar{F}_{1}\rangle }{R_{b}}.  \label{eq.2n}
\end{equation}%
Here, $\bar{F}_{1}$ is given by
\begin{equation}
\bar{F}_{1}=\frac{2T_{1}(f_{S}\cos \bar{\theta}_{L}-g_{S}\sin \bar{\theta}
_{L})}{2-T_{1}+T_{1}(\cos \bar{\theta} _{L}g_{S}+\sin \bar{\theta} _{L}f_{S})%
}
\end{equation}%
for spin-triplet superconductor and
\begin{equation}
\bar{F}_{1}=\frac{2T_{1}(-f_{S}\cos \bar{\theta} _{L}-g_{S}\sin \bar{\theta}
_{L})}{2-T_{1}+T_{1}(\cos \bar{\theta} _{L}g_{S} - \sin \bar{\theta}
_{L}f_{S})}
\end{equation}%
for spin-singlet superconductor respectively. At the N/DF interface, the boundary condition reads 
\begin{equation}
\frac{L}{R_{d}}(\frac{\partial \bar{\theta} }{\partial x})\mid _{x=0}=-\frac{%
\langle \bar{F}_{2}\rangle }{R^{\prime }_{b}}, \bar{F}_{2}=\frac{2T_{2} \sin
\bar{\theta} _{0}}{2-T_{2}+T_{2}\cos \bar{\theta} _{0}}.  \label{eq.3n}
\end{equation}%
Here $\bar{\theta} _{L}=\bar{\theta} \mid _{x=L}$ and $\bar{%
\theta} _{0}=\bar{\theta} \mid _{x=0}$.

Equations (\ref{eq.1n}) and (\ref{eq.2n}) can be transformed to
\begin{equation}
D\frac{\partial ^{2}\bar{\theta}^{\ast }(-\varepsilon )}{\partial x^{2}}%
+2i(\varepsilon +h)\sin \bar{\theta}^{\ast }(-\varepsilon )=0  \label{eq.1d}
\end{equation}%
\begin{equation}
\frac{L}{R_{d}}(\frac{\partial \bar{\theta}^{\ast }(-\varepsilon )}{\partial
x})\mid _{x=L}=\frac{\langle \bar{F}_{1}^{\ast }(-\varepsilon )\rangle }{%
R_{b}},  \label{eq.2d}
\end{equation}%
\begin{equation}
\frac{L}{R_{d}}(\frac{\partial \bar{\theta}^{\ast }(-\varepsilon )}{\partial
x})\mid _{x=0}=-\frac{\langle \bar{F}_{2}^{\ast }(-\varepsilon )\rangle }{%
R_{b}^{\prime }}.  \label{eq.3d}
\end{equation}%
The pair amplitude is defined as%
\begin{equation}
f_{3}(\varepsilon )=(\sin \theta -\sin \bar{\theta})/2
\end{equation}%
in the spin-singlet case and as
\begin{equation}
f_{0}(\varepsilon )=(\sin \theta +\sin \bar{\theta})/2
\end{equation}
in the spin-triplet case.

Since only an even-parity $s$-wave pairing can exist in the DF due to the
impurity scattering, $f_{3}$ and $f_{0}$ belong to the ESE and OTE state,
respectively.

%
In the following, we will consider four possible symmetry classes of
superconductivity in the junction, consistent with the Pauli's principle:
ESE, ETO, OTE and OSO pairing states.

(1) Junction with ESE superconductor

In this case, $f_{\pm }(\varepsilon )=f_{\pm }^{\ast }(-\varepsilon )$ and $%
g_{\pm }(\varepsilon )=g_{\pm }^{\ast }(-\varepsilon )$ are satisfied. Then,
$f_{S}(-\varepsilon )=f_{S}^{\ast }(\varepsilon )=f_{S}^{\ast }$ and $%
g_{S}(-\varepsilon )=g_{S}^{\ast }(\varepsilon )=g_{S}^{\ast }$ and we
obtain for $\bar{F}_{1}^{\ast }(-\varepsilon )$ 
\begin{equation*}
\bar{F}_{1}^{\ast }(-\varepsilon )=\frac{2T_{1}[-f_{S}\cos \bar{\theta}
_{L}^{\ast }(-\varepsilon )-g_{S}\sin \bar{\theta} _{L}^{\ast }(-\varepsilon
)]}{2-T_{1}+T_{1}[\cos \bar{\theta} _{L}^{\ast }(-\varepsilon )g_{S} -\sin
\bar{\theta}_{L}^{\ast }(-\varepsilon )f_{S}]}.
\end{equation*}%
It follows from a comparison of Eqs. \ref{eq.1}-\ref{eq.3} with Eqs. \ref%
{eq.1d}-\ref{eq.3d} that these equations are consistent with each other only
when $\sin \bar{\theta} ^{\ast }(-\varepsilon ) =-\sin {\theta}
(\varepsilon )$ and $\cos \bar{\theta} ^{\ast }(-\varepsilon )=\cos {%
\theta} (\varepsilon )$. After simple calculation, we can show $%
f_{3}(\varepsilon)=f_{3}^{*}(-\varepsilon)$ and $f_{0}(%
\varepsilon)=-f_{0}^{*}(-\varepsilon)$. This relation is consistent with the
fact \cite{Tanaka2006} that $f_{3}$ and $f_{0}$ are the even-frequency and
odd-frequency pairing state, respectively. When $h$=0, since $%
\sin\theta(\varepsilon)=-\sin\bar{\theta}(\varepsilon)$ is satisfied, the
resulting $f_{0}$ is vanishing and only the ESE state exist. For $h \neq 0$,
$f_{0}$ becomes nonzero and the OTE state is generated in DF.

(2) Junction with ETO superconductor

Now we have $f_{\pm }(\varepsilon )=f_{\pm }^{\ast }(-\varepsilon )$ and $%
g_{\pm }(\varepsilon )=g_{\pm }^{\ast }(-\varepsilon )$. Then, $%
f_{S}(-\varepsilon )=-f_{S}^{\ast }(\varepsilon )=-f_{S}^{\ast }$ and $%
g_{S}(-\varepsilon )=g_{S}^{\ast }(\varepsilon )=g_{S}^{\ast }$. As a
result, $\bar{F}_{1}^{\ast }(-\varepsilon )$ is given by

\begin{equation*}
\bar{F}_{1}^{\ast }(-\varepsilon )=-\frac{2T_{1}[f_{S}\cos \bar{\theta}%
_{L}^{\ast }(-\varepsilon )+g_{S}\sin \bar{\theta}_{L}^{\ast }(-\varepsilon
)]}{2-T_{1}+T_{1}[\cos \bar{\theta}_{L}^{\ast }(-\varepsilon )g_{S}-\sin
\bar{\theta}_{L}^{\ast }(-\varepsilon )f_{S}]}.
\end{equation*}%
Eqs. \ref{eq.1}-\ref{eq.3} and Eqs. \ref{eq.1d}-\ref{eq.3d} are consistent
if $\sin \theta ^{\ast }(-\varepsilon )=-\sin \bar{\theta}(\varepsilon )$
and $\cos \theta ^{\ast }(-\varepsilon )=\cos \bar{\theta}(\varepsilon )$.
As in the case of ESE pairing, we can show $f_{3}(\varepsilon )=f_{3}^{\ast
}(-\varepsilon )$ and $f_{0}(\varepsilon )=-f_{0}^{\ast }(-\varepsilon )$.
For $h=0$, OTE state is generated in the DF as shown in our recent paper
\cite{Tanaka2006}. The ESE state is generated by $h$, in contrast to the
case of DF/ESE superconductor junctions.

(3) Junction with OTE superconductor

In this case $f_{\pm }(\varepsilon )=-f_{\pm }^{\ast }(-\varepsilon )$ and $%
g_{\pm }(\varepsilon )=g_{\pm }^{\ast }(-\varepsilon )$. Then $%
f_{S}(-\varepsilon )=-f_{S}^{\ast }(\varepsilon )$ and $g_{S}(-\varepsilon
)=g_{S}^{\ast }(\varepsilon )$ and one can show that $\bar{F}_{1}^{\ast
}(-\varepsilon )$ has the same form as in the case of ESE and ETO
superconductor junctions. Then, we obtain $\sin \bar{\theta}^{\ast
}(-\varepsilon )=-\sin \theta (\varepsilon )$ and $\cos \bar{\theta}^{\ast
}(-\varepsilon )=\cos \theta (\varepsilon )$. Also $f_{3}(\varepsilon
)=f_{3}^{\ast }(-\varepsilon )$ and $f_{0}(\varepsilon )=-f_{0}^{\ast
}(-\varepsilon )$ are satisfied. For $h=0$, only the OTE pairing state is
generated in DF. Similar to the case of ETO junctions, ESE pairing is
induced in the presence of $h$.

(4) Junction with OSO superconductor

We have $f_{\pm }(\varepsilon )=-f_{\pm }^{\ast }(-\varepsilon )$, $g_{\pm
}(\varepsilon )=g_{\pm }^{\ast }(-\varepsilon )$,  $f_{S}(-\varepsilon
)=f_{S}^{\ast }(\varepsilon )$, and $g_{S}(-\varepsilon )=g_{S}^{\ast
}(\varepsilon )$. 
One can show that $\bar{F}_{1}^{\ast }(-\varepsilon )$ takes the same form
as in the cases of ESE, ETO, OTE superconductor junctions. 
Then, we obtain  $\sin \bar{\theta}^{\ast }(-\varepsilon )=-\sin \theta
(\varepsilon )$ and $\cos \bar{\theta}^{\ast }(-\varepsilon )=\cos \theta
(\varepsilon )$. Also $f_{3}(\varepsilon )=f_{3}^{\ast }(-\varepsilon )$ and
$f_{0}(\varepsilon )=-f_{0}^{\ast }(-\varepsilon )$ are satisfied. For $h=0$%
, only the ESE pairing state is generated in DF. Similar to the case of ETO
junctions, OTE pairing is induced in the presence of $h$.

We can now summarize the above results in the table below. As seen from the
above discussion, $\sin \bar{\theta}^{\ast }(-\varepsilon )=-\sin \theta
(\varepsilon )$, $\cos \bar{\theta}^{\ast }(-\varepsilon )=\cos \theta
(\varepsilon )$, $f_{3}(\varepsilon )=f_{3}^{\ast }(-\varepsilon )$ and $%
f_{0}(\varepsilon )=-f_{0}^{\ast }(-\varepsilon )$ are satisfied for all
cases. The real part of $f_{3}$ is an even function of $\varepsilon $ while
the imaginary part of it is an odd function of $\varepsilon $ consistent
with even-frequency pairing. On the other hand, the real part of $f_{0}$ is
an odd function of $\varepsilon $ while its imaginary part is an even
function of $\varepsilon $ consistent with odd-frequency pairing.

\begin{center}
\begin{tabular}{|c|p{2.5cm}|p{2.5cm}|p{2.5cm}|}
\hline
& Symmetry of the pairing in superconductors & Symmetry of the pairing in DF
without exchange field & Symmetry of the pairing in DF \\ \hline
(1) & Even-frequency spin-singlet even-parity (ESE) & ESE & ESE + \textbf{OTE%
} \\ \hline
(2) & Even-frequency spin-triplet odd-parity (ETO) & OTE & OTE + \textbf{ESE}
\\ \hline
(3) & Odd-frequency spin-triplet even-parity (OTE) & OTE & OTE + \textbf{ESE}
\\ \hline
(4) & Odd-frequency spin-singlet odd-parity (OSO) & ESE & ESE + \textbf{OTE}
\\ \hline
\end{tabular}
\end{center}

Within this formulation, the LDOS in the DF layer is given by
\begin{equation}
N/N_{0}=\frac{1}{2}(\mathrm{Re}\cos \theta +\mathrm{Re}\cos \bar{\theta})
\end{equation}%
where $N_{0}$ denotes the LDOS in the normal state. Below we will calculate $%
f_{3}$, $f_{0}$ and LDOS at zero temperature. For this purpose, we will use
the following parameter set $Z=3$, $Z^{\prime }=3$, $E_{Th}\equiv
D/L^{2}=0.1\Delta $ and $R_{d}/R_{b}^{\prime }=0.1$, which represents a
typical DF/S junction. Our qualitative conclusions are not sensitive to the
parameter choice.



\section{Results}

In the following, we will study two typical cases. As an example of ESE
superconductor, the conventional spin-singlet $s$-wave pairing will be
considered. We will clarify the generation of OTE pairing in DF by the
exchange field $h$ consistent with preexisting results \cite{Efetov1,Efetov2}%
. We will also study spin-triplet $p$-wave superconductor as a typical
example of ETO superconductor. In this case, ESE pairing state is induced by
$h$. It should be remarked again that $f_{3}$ and $f_{0}$ denote the ESE and
OTE pairing amplitudes, respectively.

\subsection{Spin singlet $s$-wave superconductor junctions}

\begin{figure}[htb]
\begin{center}
\scalebox{0.4}{
\includegraphics[width=22.0cm,clip]{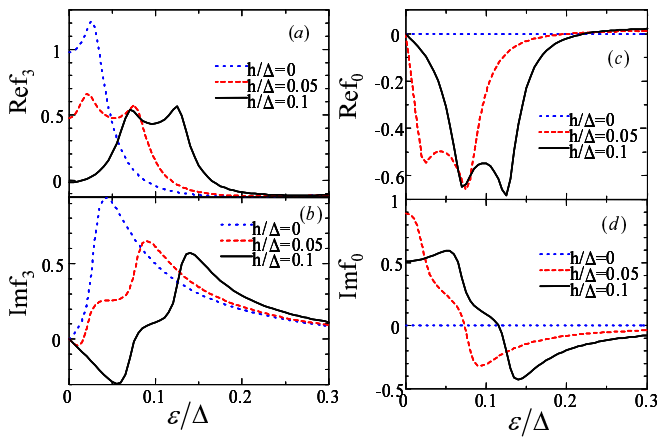}}
\end{center}
\caption{ (color online) Real (a) and imaginary (b) parts of $f_3$, and real
(c) and imaginary (d) parts of $f_0$ in spin-singlet $s$-wave superconductor
junctions. We choose $R_d/R_b = 1$. }
\label{f2}
\end{figure}

\begin{figure}[tbh]
\begin{center}
\scalebox{0.4}{
\includegraphics[width=18.0cm,clip]{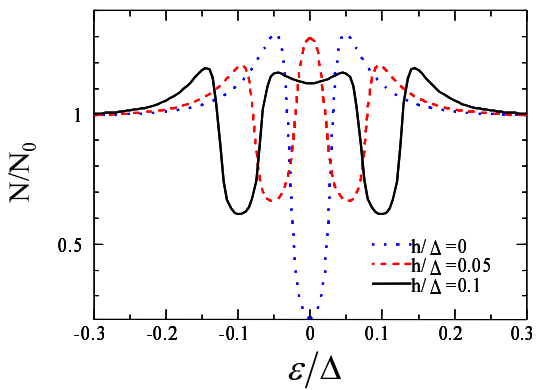}}
\end{center}
\caption{ (color online) Normalized LDOS as a function of $\protect%
\varepsilon $ for $R_{d}/R_{b}=1$ with various $h/\Delta $ in spin-singlet $s
$-wave superconductor junctions.}
\label{f3}
\end{figure}

Let us first study DF/spin-singlet $s$-wave superconductor junctions where
we choose $R_d/R_b = 1$ and the form factor $\Psi_{\pm}$ is given by $%
\Psi_{\pm}=1$. Real and imaginary parts of $f_3$ and $f_0$ at $x=0$ for various $%
h/\Delta$ are shown in Fig. \ref{f2}. Without exchange field, $i.e.$, $h=0$,
only the $f_{3}$ is nonzero, consistent with conventional theory of
proximity effect \cite{Volkov,Yip,TGK}.
By introducing the exchange field $h$, the magnitude of $f_{3}$ is
suppressed for small $\varepsilon$ while it is enhanced for large $%
\varepsilon$ as shown in Figs. \ref{f2}(a) and \ref{f2}(b). On the other
hand, the imaginary part of $f_{0}$ is enhanced for small magnitude of $%
\varepsilon$. The corresponding LDOS at N/DF interface normalized by its
value in the normal state is plotted as a function of $\varepsilon$ in Fig. %
\ref{f3}. The LDOS has a minigap at $h=0$ \cite{Volkov,Yip}. As shown in
Fig. \ref{f3}, the LDOS is influenced crucially by $h$. A peak appears at
zero energy with $h/\Delta=0.05$. In this case Im$f_0$ has a large value at
zero energy as shown in Fig. \ref{f2}(d). Thus large magnitude of Im$f_0$ at
$\varepsilon=0$ is responsible for the peak of the LDOS.

\begin{figure}[htb]
\begin{center}
\scalebox{0.4}{
\includegraphics[width=15.0cm,clip]{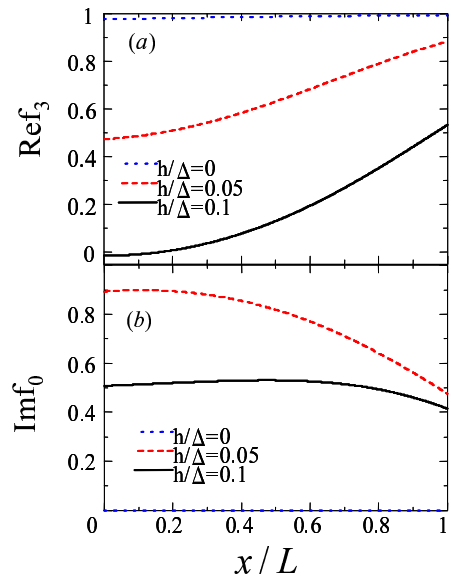}}
\end{center}
\caption{ (color online) Spatial dependence of the pair amplitudes $f_{3}$
and $f_{0}$ in DF for $\protect\varepsilon=0$ in spin-singlet $s$-wave
superconductor junctions. For $\protect\varepsilon=0$, $\mathrm{Im}f_{3}=0$
and $\mathrm{Re}f_{0}=0$ are satisfied. }
\label{f4}
\end{figure}

It was shown in our previous work \cite{Yoko} that the condition for the
formation of ZEP in the LDOS is given by $E_{Th}\sim 2hR_{b}/R_{d}$. This
condition is consistent with the results shown in Fig. 3.
As shown in Fig. \ref{f2}, when this condition is satisfied, $\mathrm{Im}%
f_{0}$ has a large value at the zero energy. Thus it corresponds to the
generation of the odd-frequency pairing amplitude $f_{0}$ at low energy. The
spatial dependences of the pair amplitudes $f_{3}$ and $f_{0}$ at $%
\varepsilon =0$ are shown in Fig. \ref{f4}. The amplitude of $f_{3}$ is
dominant near the DF/S interface while the magnitude of $f_{0}$ is enhanced
at the N/DF interface.

\begin{figure}[htb]
\begin{center}
\scalebox{0.4}{
\includegraphics[width=18.0cm,clip]{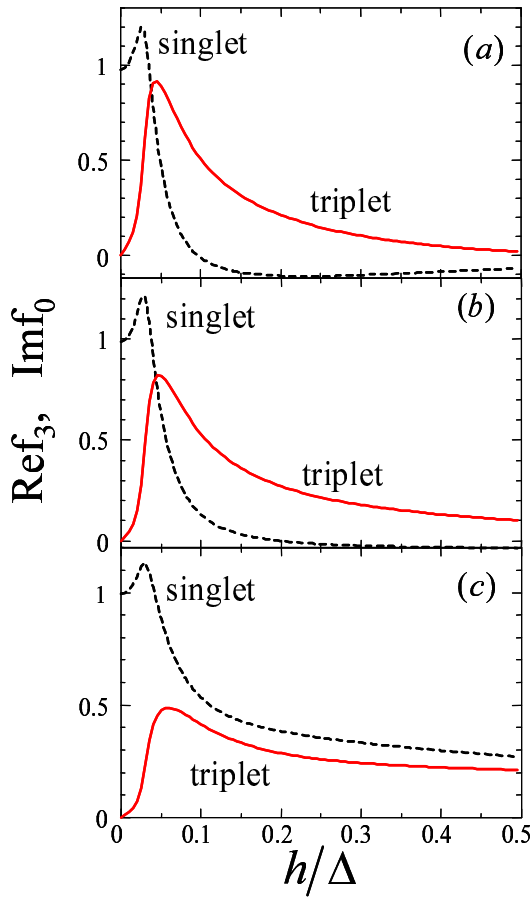}}
\end{center}
\caption{ (color online) The pair amplitudes $f_{3}$ and $f_{0}$ as a
function of $h$ in DF for $\protect\varepsilon=0$ in spin-singlet $s$-wave
superconductor junctions. (a) $x=0$. (b) $x=L/2$. (c) $x=L$. }
\label{f5}
\end{figure}

Let us study the crossover between singlet and triplet pairing states. We
show $f_{3}$ and $f_{0}$ as a function of $h$ for $\varepsilon =0$ at (a) $%
x=0$, (b) $x=L/2$ and (c) $x=L$ in Fig. \ref{f5}. $f_{0}$ increases from
zero with $h$. At a certain value of $h$, $f_{0}$ has a maximum. If the
value of $h$ is larger than this value, the triplet component becomes
dominant as shown in Fig. \ref{f5}(a) and Fig. \ref{f5}(b). The value of $h$
at the crossover regime is given by the minigap in DN/S junctions. Let us
discuss this regime in more detail. As shown in section II, $\sin \bar{\theta%
}(\varepsilon )=-\sin \theta ^{\ast }(-\varepsilon )$ and $\cos \bar{\theta}%
(\varepsilon )=\cos \theta ^{\ast }(-\varepsilon )$ are satisfied for any
case. Then the ESE and OTE pair wave functions in the DF are given by
\begin{eqnarray}
f_{3}(\varepsilon ) &=&[\sin \theta (\varepsilon )+\sin \theta ^{\ast
}(-\varepsilon )]/2, \\
f_{0}(\varepsilon ) &=&[\sin \theta (\varepsilon )-\sin \theta ^{\ast
}(-\varepsilon )]/2.
\end{eqnarray}%
At $\varepsilon =0$, we denote $\theta (0)=\mathrm{Re}\theta (0)+i\mathrm{Im}%
\theta (0)$, where $\mathrm{Re}\theta (0)$ and $\mathrm{Im}\theta (0)$ are
the real and imaginary part of $\theta (0)$. Then $f_{3}(0)$ and $f_{0}(0)$
are given by $\cosh [\mathrm{Im}\theta (0)]\sin [
\mathrm{Re}\theta (0)]$ and $i\sinh [\mathrm{Im}\theta (0)]\cos [\mathrm{Re}\theta (0)]$. Thus the following equation is satisfied:
\begin{equation}
\frac{{f_{3}(0)}}{{f_{0}(0)}}=\frac{{\tan {\mathop{\rm Re}\nolimits}\theta
(0)}}{{i\tanh {\mathop{\rm Im}\nolimits}\theta (0)}}.
\end{equation}%
It is easy to show that $\left\vert {{\mathop{\rm Re}\nolimits}\theta (0)}%
\right\vert <\left\vert {{\mathop{\rm Im}\nolimits}\theta (0)}\right\vert $
is satisfied when the crossover occurs, i.e., $\tan {\mathop{\rm Re}\nolimits%
}\theta (0)=\tanh {\mathop{\rm Im}\nolimits}\theta (0)$. As shown in our
previous work\cite{Yoko}, this inequality is satisfied when the exchange
field is of the order of the minigap energy in DN/S junctions, i.e., $h\sim
(R_{d}/R_{b})(E_{Th}/2)$. Therefore the crossover occurs around this value
of the exchange field.

\subsection{Spin-triplet $p$-wave superconductor junctions}

\begin{figure}[tbh]
\begin{center}
\scalebox{0.4}{
\includegraphics[width=23.0cm,clip]{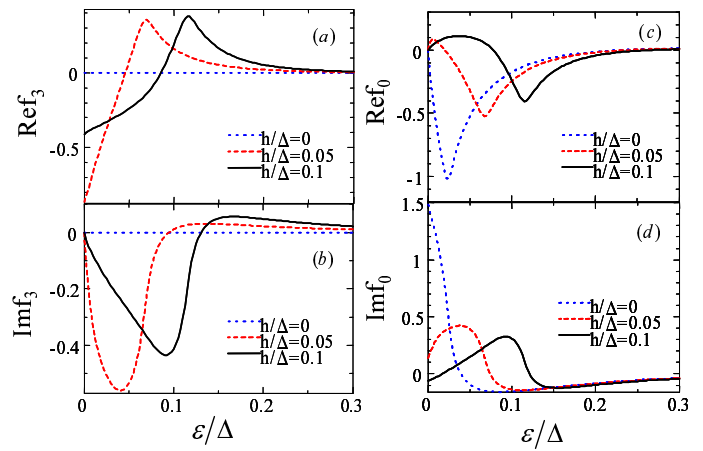}}
\end{center}
\caption{ (color online)Pair amplitudes for DF/ spin-triplet $p$-wave
superconductor junctions. Real (a) and imaginary (b) parts of $f_{3}$. Real
(c) and imaginary (d) parts of $f_{0}$. Here we choose $R_{d}/R_{b}=0.1$.}
\label{f6}
\end{figure}

Next we focus on the DF / spin-triplet $p$-wave superconductor junctions,
where we choose $R_{d}/R_{b}=0.1$ and the form factor $\Psi _{\pm }$ is
given by $\Psi _{\pm }=\pm \cos \phi $ corresponding to the case of $\alpha
=0$ (see Fig. \ref{f1}). In order to make numerical calculations stable, we
introduce small imaginary number in the quasiparticle energy: $\varepsilon
\rightarrow \varepsilon +i\gamma $, with $\gamma =0.01\Delta $. The real and
imaginary parts of $f_{3}$ and $f_{0}$ at $x=0$ are plotted in Fig. \ref{f6} for various $h/\Delta $. Similar to the case of DN/s-wave superconductor
junctions, the imaginary part of $f_{3}$ and the real part of $f_{0}$ vanish
at $\varepsilon =0$. For $h$=0, $f_{3}=0$ and only $f_{0}$ is nonzero as
shown in Fig. \ref{f6}.
The feature of this unusual proximity effect\cite{p-wave} was already
discussed in our previous paper\cite{Tanaka2006}, where OTE pairing state is
generated in the DN of DN/ETO superconductor junctions. In this case, the
LDOS has a ZEP and odd-frequency component $f_{0}$ becomes a purely
imaginary number at $\varepsilon =0$.
\begin{figure}[tbh]
\begin{center}
\scalebox{0.4}{
\includegraphics[width=18.0cm,clip]{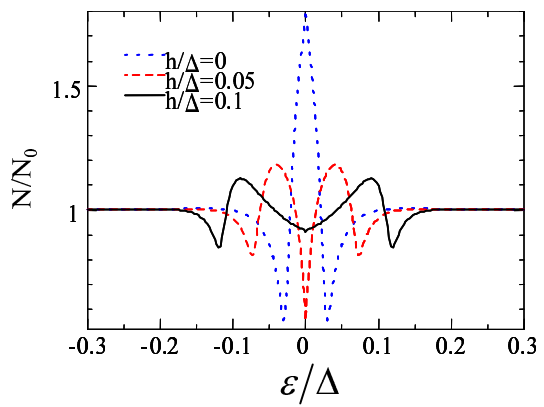}}
\end{center}
\caption{(color online) Normalized LDOS as a function of $\protect%
\varepsilon $ for $R_{d}/R_{b}=0.1$ and various $h/\Delta $ in $p$-wave
superconductor junctions.}
\label{f7}
\end{figure}
With increasing $h$, the amplitude of $f_{3}$ is enhanced as shown in Figs. %
\ref{f6}(a) and \ref{f6}(b), in contrast to the case of DN/spin-singlet
s-wave superconductor junctions. At the same time, the magnitude of $f_{0}$
near the zero energy is suppressed. Then the features of the proximity
effect in DF are the same as in conventional superconductor junctions. The
corresponding LDOS normalized by its value in the normal state is plotted as
a function of $\varepsilon $ in Fig. \ref{f7}. With the increase of $h$, the
magnitude of LDOS at $\varepsilon =0$ is suppressed and the LDOS peak is
splitted. The magnitude of the splitting increases with the increase of $h$.
Note that the peak positions in Im$f_{0}$ and LDOS coincide with each other.
\begin{figure}[tbh]
\begin{center}
\scalebox{0.4}{
\includegraphics[width=16.0cm,clip]{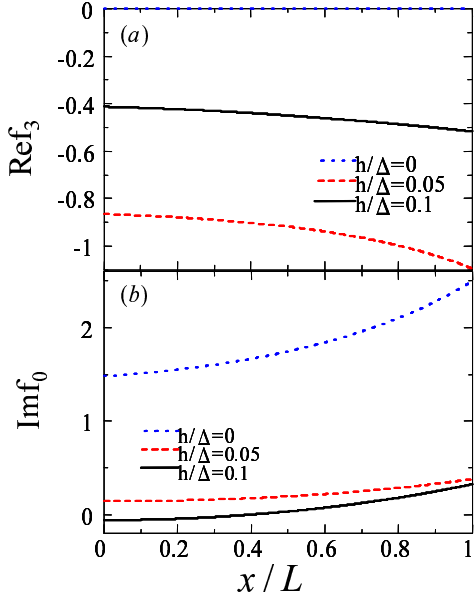}}
\end{center}
\caption{(color online) Spatial dependence of the pair amplitudes $f_{3}$
and $f_{0}$ in DF for $\protect\varepsilon =0$ in $p$-wave superconductor
junctions. For $\protect\varepsilon =0$, $\mathrm{Im}f_{3}=0$ and $\mathrm{Re%
}f_{0}=0$ are satisfied.}
\label{f8}
\end{figure}
The spatial dependences of the real part of $f_{3}$ and the imaginary part
of $f_{0}$ at $\varepsilon =0$ are shown in Fig. \ref{f8}. For $h=0$, $f_{3}$
is absent and the magnitude of the imaginary part of $f_{0}$ reaches its
maximum at the DF/S interface. With the increase of $h$, the amplitude of $%
f_{0}$ is drastically reduced. The spatial dependence of $f_{3}$ is rather
weak and its amplitude is most strongly enhanced for $h=0.05\Delta $. At the
same time, the magnitude of LDOS at $\varepsilon =0$ is most strongly
suppressed (see Fig. \ref{f7}).

\begin{figure}[htb]
\begin{center}
\scalebox{0.4}{
\includegraphics[width=18.0cm,clip]{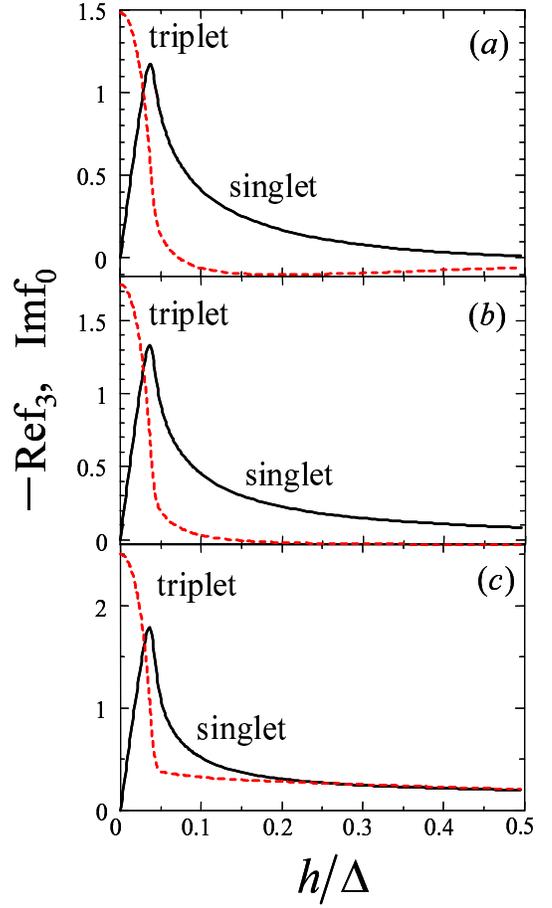}}
\end{center}
\caption{ (color online) The pair amplitudes $f_{0}$ and $f_{3}$ as a
function of $h$ in DF for $\protect\varepsilon=0$ in $p$-wave superconductor
junctions. (a) $x=0$. (b) $x=L/2$. (c) $x=L$. }
\label{f9}
\end{figure}

Before ending this subsection, we investigate the crossover between singlet
and triplet pairing states. Let us plot $f_{3}$ and $f_{0}$ for $\varepsilon
=0$ as a function of $h$ at (a) $x=0$, (b) $x=L/2$ and (c) $x=L$ in Fig. \ref%
{f9}. $f_{3}$ has a maximum at a certain value of $h$. When $h$ exceeds this
value, the singlet component becomes dominant as shown in Fig. \ref{f9}. The
value of $h$ at the crossover increases with the increase of $Z$, $%
R_{d}/R_{b}$ and $E_{Th}$, i.e., with the enhancement of the proximity
effect.

\subsection{Relevance of the odd-frequency component to ZEP of LDOS}

Let us discuss the relation between the generation of the odd-frequency
pairing and ZEP in LDOS, using general properties of solutions of the
proximity effect problem. Since $\cos \bar{\theta}(\varepsilon )=\cos \theta
^{\ast }(-\varepsilon )$ are satisfied, the LDOS normalized by its value in
the normal state is given by
\begin{equation}
N/N_{0}=[\cos \theta (\varepsilon )+\cos \theta ^{\ast }(-\varepsilon )]/2.
\end{equation}

For $\varepsilon =0$, the normalized LDOS reads $\cosh [\mathrm{Im}\theta
(0)]\cos [\mathrm{Re}\theta (0)]$, while $f_{3}(0)$ and
$f_{0}(0)$ are given by $\cosh [\mathrm{Im}\theta (0)]\sin [\mathrm{Re}\theta (0)]$ and $i\sinh [\mathrm{Im}\theta (0)]\cos [\mathrm{Re}\theta (0)]$ respectively. As seen from
these relations, $f_{0}$ becomes zero when the LDOS is zero. In addition,
whether the spin-singlet component $f_{3}$ dominates the spin-triplet
component $f_{0}$ or not crucially depends on the value of $\mathrm{{Re}%
\theta (0)}$. The most favorable condition where $N/N_{0}$ is enhanced is
the large magnitude of $\mathrm{Im}\theta (0)$ and the absence of $\mathrm{Re%
}\theta (0)$, where $f_{0}$ dominates $f_{3}$. %
%
For the sufficiently large magnitude of $\mathrm{Im}\theta (0)$ and small
magnitude of $\mathrm{Re}\theta (0)$, $N/N_{0}\sim \cos [\mathrm{Re}\theta
(0)]\exp [\mathrm{Im}\theta (0)]/2\sim \exp [\mathrm{Im}\theta
(0)]/2$ and $f_{0}(0)\sim i\cos [\mathrm{Re}\theta (0)]\exp [\mathrm{Im}\theta (0)]/2\sim i\exp [\mathrm{Im}\theta (0)]/2$ are satisfied.
Then we obtain $N/N_{0}\sim -if_{0}(0)$. This means that the generation of
the odd-frequency pair amplitude $f_{0}(0)$ leads to the enhancement of the
density of states at zero energy.


\section{Conclusions}

We have studied the proximity effect in diffusive ferromagnet (DF) /
superconductor (S) junctions. Various possible symmetry classes in a
superconductor were considered which are consistent with the Pauli's
principle: even-frequency spin-singlet even-parity (ESE) state,
even-frequency spin-triplet odd-parity (ETO) state, odd-frequency
spin-triplet even-parity (OTE) state and odd-frequency spin-singlet
odd-parity (OSO) state. As was established in the previous work \cite%
{Tanaka2006}, in the absence of the exchange field the induced pair
amplitude in a DF belongs respectively to ESE, OTE, OTE and ESE pairing
states. It is shown in the present paper that, in addition to these states,
the OTE, ESE, ESE and OTE pairing states are generated in DF in the presence
of the exchange field $h$.

As a typical example of ESE superconductor, we have chosen spin-singlet $s$-wave state. We have clarified that when the OTE state dominates the ESE
state in the DF, the resulting LDOS has a zero energy peak. At the same time, 
the amplitude of the OTE pair wave function near the N/DF interface is enhanced  at zero energy. As suggested by our findings, the odd-frequency pairing state was possibly realized in the experiment by Kontos\cite{Kontos}, where the ZEP was observed in ferromagnet / $s$-wave superconductor junctions.

We have also studied spin-triplet $p$-wave superconductor junctions. In this
case, the ZEP in the LDOS splits into two peaks due to the generation of the
ESE pairing state by the exchange field. The features of proximity effect
specific to spin-triplet $p$-wave superconductor junctions can be studied in
experiments with Sr$_{2}$RuO$_{4}$-Sr$_{3}$Ru$_{2}$O$_{7}$ eutectic system
\cite{Maeno1}. Based on general properties of solutions of the proximity
effect problem, we have demonstrated that the generation of the
odd-frequency pairing state at zero energy leads to the ZEP in LDOS. %

The authors appreciate useful and fruitful discussions with M. Eschrig and
Ya. V. Fominov. T. Y. acknowledges support by JSPS Research Fellowships for
Young Scientists. This work was supported by Grant-in-Aid for Scientific
Research on Priority Area "Novel Quantum Phenomena Specific to Anisotropic
Superconductivity" (Grant No. 17071007) from the Ministry of Education,
Culture, Sports, Science and Technology of Japan. This work was also
supported by NAREGI Nanoscience Project, the Ministry of Education, Culture,
Sports, Science and Technology, Japan, the Core Research for Evolutional
Science and Technology (CREST) of the Japan Science and Technology
Corporation (JST) and a Grant-in-Aid for the 21st Century COE "Frontiers of
Computational Science" . The computational aspect of this work has been
performed at the Research Center for Computational Science, Okazaki National
Research Institutes and the facilities of the Supercomputer Center,
Institute for Solid State Physics, University of Tokyo.

%

\end{document}